\documentclass[aps,prc,twocolumn,amsmath,superscriptaddress,floatfix,nofootinbib]{revtex4-1}

\usepackage{setspace,ulem, epsfig,amssymb,amsfonts,amsmath,mathtools,bm,color,xcolor,graphicx,braket,adjustbox,esint,upgreek,subcaption,verbatim,ctable}

\usepackage{multirow}
\usepackage{array}
\usepackage{ragged2e}
\usepackage{graphicx}
\graphicspath{
    {./Fig/} 
}
\usepackage[colorlinks,linkcolor=red,anchorcolor=green,citecolor=blue]{hyperref}

\newcommand{\tred}{\textcolor{black}}

\begin{document}
%\begin{spacing}{1.0}
\title{Physics-Informed Neural Network for Solving the Heavy quark diffusion in the Expanding QCD Medium}

\author{Wenhua Fan}
\affiliation{The International Joint Institute of Tianjin University, Fuzhou,  Tianjin University, Tianjin 300072, China}

\author{Jiamin Liu}
\affiliation{Department of Physics, School of Science, Tianjin University, Tianjin 300354, China}

\author{Huansang Yang}
\affiliation{College of Engineering, Georgia Institute of Technology, Atlanta, GA 30332, United States}

\author{Baoyi Chen}
\email{baoyi.chen@tju.edu.cn}
\affiliation{The International Joint Institute of Tianjin University, Fuzhou,  Tianjin University, Tianjin 300072, China}
\affiliation{Department of Physics, School of Science, Tianjin University, Tianjin 300354, China}

\date{\today}

\begin{abstract}
\tred{We employ Physics-Informed Neural Networks (PINNs) to investigate the dynamical evolution of heavy quarks within the expanding hot QCD medium generated in relativistic heavy-ion collisions. The heavy quark dynamics are first modeled under the assumption of complete kinetic thermalization, followed by a more realistic study of non-thermal diffusion governed by the Fokker-Planck (FP) equation. In both scenarios, the background evolution of the hot QCD medium is encoded into the coefficients of the diffusion equations. These equations are solved within the PINN framework, where the initial conditions, physical constraints from the dynamical equation, and probability conservation are incorporated into the loss function.We also compare the performance of the FP-PINN with a supervised five-dimensional DNN trained on labeled data generated from Langevin-based reference distributions.This work provides a valuable reference for applying PINN-based models to particle diffusion in phase space, laying the foundation for future studies of heavy quarkonium production via realistic non-thermal heavy-quark coalescence.
}

\end{abstract}
%\pacs{ }
\maketitle

\section{Introduction}

Relativistic heavy-ion collisions produce a strongly coupled, deconfined medium known as the quark-gluon plasma (QGP)~\cite{Bazavov:2011nk}. Over the past several decades, experiments at the Relativistic Heavy Ion Collider (RHIC) and the Large Hadron Collider (LHC) have measured the production and momentum distributions of various hadrons to characterize the properties of strong interactions at extreme temperatures~\cite{PHENIX:2003iij,ALICE:2010suc}. Extensive studies interpreting these data indicate that the deconfined medium behaves as a nearly perfect fluid~\cite{Pang:2012he,Gale:2013da,Heinz:2013th}. Heavy quarks, produced via hard parton scatterings at the onset of nuclear collisions, serve as sensitive probes of this early-stage medium, which typically possesses a lifetime on the order of a few ${\rm fm}/c$~\cite{Matsui:1986dk, He:2012hq, He:2013ds, Liu:2010ej}. The dynamical evolution of heavy quarks is commonly described by the Langevin equation~\cite{Cao:2015hia,Chen:2021akx} or the Boltzmann-type transport equation~\cite{Xing:2024qcr}. Due to their strong coupling with the bulk medium, heavy quarks undergo significant energy loss in Pb-Pb collisions at LHC energies. In collisions where multiple charm-quark pairs are produced, charm and anti-charm quarks may recombine within the deconfined medium, hadronizing into bound-state heavy quarkonia. This process, known as regeneration or coalescence, can significantly enhance the charmonium yield—a phenomenon well-established by experimental observations~\cite{Thews:2000rj,Greco:2003vf,Andronic:2003zv,Yan:2006ve}. Because the charmonium yield from coalescence is highly sensitive to the phase-space density of charm quarks~\cite{Zhou:2014kka,Pan:2023ouw}, accurately determining this density within the expanding hot QCD medium is essential for a quantitative understanding of both open and hidden heavy-flavor production.

\tred{In the strong-coupling limit, the momentum distribution of heavy quarks satisfies local kinetic equilibrium, while their spatial distribution satisfies the diffusion equation. A more complete nonequilibrium description is provided by the Fokker-Planck equation in phase space~\cite{He:2012hq,Cao:2015hia}. The transport of heavy quarks is closely related to the evolution of the bulk medium~\cite{Singh:2023smw}.}
The Physics-Informed Neural Network (PINN) framework was first introduced by Raissi et al.~\cite{RAISSI2019686}. PINNs transform the task of solving partial differential equations (PDEs) into an optimization problem by minimizing a loss function. This loss function is defined by the residuals of the governing PDEs, alongside the initial and boundary conditions evaluated at collocation points, sampled randomly or systematically, within the physical domain~\cite{ai5030074}. Consequently, the underlying physical laws are encoded directly into the neural network's training process through this loss formulation.

Compared with traditional numerical methods, PINNs utilize automatic differentiation to compute derivatives efficiently. This approach eliminates the need for manual derivation and mitigates numerical errors typically associated with discretization in conventional solvers—a feature that is particularly advantageous for high-dimensional problems.
Because PINNs are mesh-free, they can effectively solve PDEs involving complex geometries or irregular boundaries without the computational overhead of mesh generation. 
\tred{PINNs have also been explored in a variety of high-dimensional and multiscale problems. In addition, some studies have employed the neural tangent kernel (NTK) framework to analyze the training dynamics and stability of PINNs~\cite{WANG2022110768}.}

\tred{PINNs have been further extended to solve stochastic differential equations~\cite{PINNsto-ref,PINNsto-ref-2} and to simulate computational fluid dynamics within complex geometries~\cite{Sedykh:2023zyh,citation-key,extended-PINN-ref}. In this work, we employ the PINN framework to investigate the transport of heavy quarks in the expanding QCD medium produced in relativistic nuclear collisions. The present approach is designed to solve the high-dimensional transport equation for a prescribed hydrodynamic background velocity field. We first consider the diffusion equation as an approximate coordinate-space description in the strong-coupling limit, and then apply the PINN model to the Fokker-Planck equation in phase space. Such a development is useful for quantitative studies of heavy-quark transport in the hot medium and provides a basis for future applications to charmonium production and related observables~\cite{Song:2017wtw,Zhao:2021voa}. In addition, to clarify the methodological distinction from purely supervised regression, we compare the phase-space PINN with a standard five-dimensional DNN trained on labeled reference data generated from Langevin equation.}

\tred{The remainder of this paper is organized as follows. We first describe the hydrodynamic evolution of the bulk medium. We then present the charm-quark diffusion equation with the background flow field incorporated, together with its further generalization to the Fokker-Planck description in phase space. Finally, we introduce the corresponding PINN framework and present the numerical results.}

\section{Charm transport equations in the hot medium}

The dynamical evolution of the strongly coupled, deconfined medium produced in relativistic heavy-ion collisions has been extensively investigated using hydrodynamic models~\cite{Song:2007ux}. In this work, we utilize the $(2+1)$-dimensional MUSIC package~\cite{Schenke:2010nt, Schenke:2010rr} to simulate the temperature and velocity profiles of the fluid in Pb-Pb collisions at $\sqrt{s_{NN}} = 5.02$ TeV. The equation of state (EoS) for the deconfined phase is derived from lattice QCD calculations, whereas the EoS for the hadronic phase is based on the Hadron Resonance Gas model~\cite{HotQCD:2014kol, Bernhard:2016tnd}. The transition between the confined and deconfined phases is modeled as a smooth crossover. Furthermore, the initial energy density of the medium is determined via the Glauber model, with additional constraints provided by the final-state charged-particle multiplicity.

\begin{figure*}[!hbt]
\centering
\includegraphics[width=1.05\textwidth]{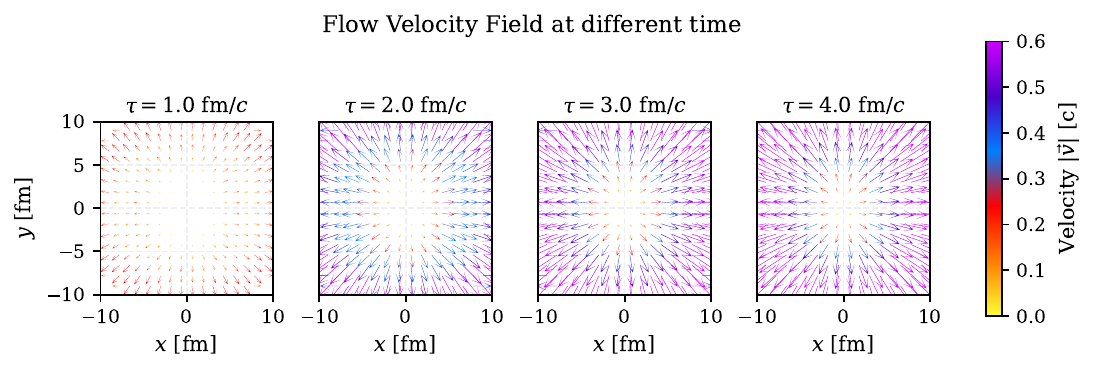}
\caption{\RaggedRight The fluid velocity distributions at the central rapidity region for Pb-Pb collisions at \mbox{$\sqrt{s_{NN}} = 5.02$} TeV generated by the (2+1) dimensional MUSIC package, with an impact parameter of \mbox{$b = 6$} fm, are presented. The direction and color of the arrows represent the local velocity vectors and their corresponding magnitudes, respectively. The four panels illustrate the temporal evolution of the velocity distribution at distinct proper time points. 
}
%\hspace{-0.1mm}
\label{lab-fig-velo}
\end{figure*}

Heavy quarks are produced during the early stages of the collision. Due to their large masses, thermal production in the hot QCD medium is negligible, and the total number of charm quarks is conserved during the medium's expansion. Lattice QCD calculations suggest a very strong coupling between heavy quarks and the hot QCD medium~\cite{Banerjee:2011ra,Francis:2015daa}. 

Assuming kinetic thermalization, the momentum distribution of charm quarks follows the normalized Fermi–Dirac distribution, while the local density $\rho_c(\tau, \mathbf{r})$ is governed by the conservation equation~\cite{Zhao:2017yan}:\begin{equation}
\label{conser}
\partial_{\mu}(\rho_c u^\mu)=0,
\end{equation}
where $u^\mu = \gamma(1, \mathbf{v})$ is the fluid four-velocity, $\gamma$ is the Lorentz factor, and $\mathbf{v}$ is the three-dimensional fluid velocity. In Pb-Pb collisions, the longitudinal expansion of the medium is effectively described by the Bjorken model. Under the assumption of full kinetic thermalization, the charm-quark density can be factorized as
$\rho_c(\tau, \mathbf{x}_T, \eta) = \rho_T(\tau, \mathbf{x}_T) \rho_\eta(\eta) \cosh\eta/\tau$,
where $\rho_T$ is the density in the transverse plane and $\eta$ is the spacetime pseudo-rapidity. The function $\rho_\eta(\eta)$ represents the normalized longitudinal distribution of charm quarks, which is often approximated as unity in the mid-rapidity region. Here, $\tau = \sqrt{t^2 - z^2}$ denotes the proper time. The longitudinal Bjorken expansion reduces the charm-quark density by a factor of $1/\tau$. In the central rapidity region, the conservation equation simplifies to:
\begin{equation}
\label{lab-eq-diff}
\frac{\partial \rho_T(\tau, x, y)}{\partial\tau}
=
-\frac{\partial(\rho_T v_x)}{\partial x}
-\frac{\partial(\rho_T v_y)}{\partial y}.
\end{equation}
where $v_x(\tau, x, y)$ and $v_y(\tau, x, y)$ represent the local transverse velocities of the medium, as provided by the hydrodynamic model. The local density of charm quarks is highly sensitive to these fluid velocity profiles. \tred{The velocity fields of the medium are considered as spatial and time-dependent coefficients in the above differential equation.} While several numerical methods, such as the finite element and finite difference methods, are available to solve this equation, they often suffer from substantial computational costs or introduce significant numerical uncertainties due to discretization. 
\tred{Physics-Informed Neural Networks can be employed to solve the partial differential equation even when the coefficients of the PDE, such as the velocity fields $v_{x,y}(\tau,x,y)$, exhibit high complexity. By avoiding the grid-based constraints of traditional numerical methods and the data-intensive requirements of purely data-driven models, this approach provides an alternative physics-informed framework for solving transport problems in heavy-quark dynamics.}

Regarding the initial spatial density of charm quarks, since they are produced via binary nucleon-nucleon collisions, their density is proportional to the product of the nuclear thickness functions:
\begin{align}
\rho_T(\tau=0, \mathbf{x}_T) = \sigma_{NN}^{c\bar{c}} T_A(\mathbf{x}_T - \mathbf{b}/2) T_B(\mathbf{x}_T + \mathbf{b}/2),
\end{align}
where $\sigma_{NN}^{c\bar{c}}$ represents the charm-pair production cross section within a specified rapidity interval. In this work, as we focus on the longitudinal and transverse diffusion of charm quarks while assuming total number conservation, this cross section is taken as unity to determine the relative charm-quark density within the expanding medium. The thickness functions $T_{A,B}$ for the lead (Pb) nuclei are calculated assuming a nucleon density that follows a Woods-Saxon distribution. The impact parameter $\mathbf{b}$ denotes the transverse distance between the centers of the two nuclei at the moment of impact. The hydrodynamic evolution of the medium, simulated for Pb-Pb collisions at $\sqrt{s_{NN}} = 5.02$ TeV, begins at an initial proper time of $\tau_0 = 0.6$ fm/$c$. We assume that for $\tau < \tau_0$, the charm quarks undergo primarily longitudinal expansion.

\tred{In the diffusion description, the momentum degrees of freedom are not treated explicitly, so that the equation effectively describes only the transport of the local density driven by the background flow field. In this reduced description, the momentum degrees of freedom are assumed to be in complete kinetic equilibrium, so that they can be integrated out. By contrast, in the Fokker-Planck description they are retained explicitly, which allows one to study the detailed thermalization process of heavy quarks in heavy ion collisions. For a more realistic description of heavy quark dynamics in the hot medium, one needs to explicitly include the momentum variables, together with the medium-induced drag and momentum-diffusion effects~\cite{He:2012hq,Cao:2015hia}. The corresponding evolution is then governed by the Fokker-Planck equation,}

\tred{
\begin{equation}
\begin{aligned}
\partial_\tau f
+\frac{p_x}{m}\partial_x f
+\frac{p_y}{m}\partial_y f
&=
\partial_{p_x}\!\Big[\gamma\,(p_x-mv_x)\,f\Big] \\
&\quad+
\partial_{p_y}\!\Big[\gamma\,(p_y-mv_y)\,f\Big] \\
&\quad+
\frac{\kappa}{2}
\left(\partial_{p_x}^2 f+\partial_{p_y}^2 f\right).
\end{aligned}
\label{eq:fp_eq}
\end{equation}}
\tred{Here $m$ is the charm-quark mass, while $\gamma$ and $\kappa$ denote the drag and momentum-diffusion coefficients, respectively. Since the drag coefficient depends on the heavy quark momentum, the corresponding momentum derivatives of $\gamma$ are automatically included in this form of the transport equation and in the numerical implementation. In the present calculation, the local temperature field $T(\tau,x,y)$ and the background flow velocities $v_x(\tau,x,y)$ and $v_y(\tau,x,y)$ are taken from the hydrodynamic background generated by MUSIC package and are treated as prescribed coefficient functions in the transport equation. Following phenomenological heavy-quark transport frameworks developed in Refs.~\cite{He:2012hq,He:2011qa,Cao:2015hia}, we parametrize the momentum-diffusion coefficient as}
\begin{equation}
\tred{
\kappa(\tau,x,y)=\frac{2\,T^2(\tau,x,y)\,[2\pi T(\tau,x,y)]}{C},}
\end{equation}
\tred{where $C$ controls the effective spatial diffusion strength. In the present work, we take $C=3$ in the QGP phase for $T>T_c$ and $C=5$ in the hadronic phase for $T\le T_c$, with $T_c=0.16$ GeV, so as to reflect the different transport properties in the two phases~\cite{He:2011qa,Cao:2015hia}. The drag coefficient is then determined through}
\begin{equation}
\gamma(\tau,x,y,p_x,p_y)
=\frac{\kappa(\tau,x,y)}{2\,T(\tau,x,y)\,E_T},
\end{equation}
where
\[
E_T=\sqrt{m^2+p_x^2+p_y^2}.
\]
\tred{so that $\gamma$ depends on both the local medium properties and the charm-quark momentum. In the present phenomenological setup, the drag and momentum-diffusion coefficients are related through an equilibrium-motivated Einstein relation, while the distribution function itself is allowed to evolve out of equilibrium. Below $T_c$, the present treatment should be understood as an effective continuation of heavy-flavor transport with modified transport coefficients, rather than a microscopic description of explicit hadronic species. In this extension, the initial spatial distribution is chosen to be the same as that in the diffusion equation, while the momentum-space part of the initial condition is specified by a normalized distribution $g_0(p_x,p_y)$ constructed from the FONLL charm-quark transverse-momentum spectrum under the assumption of azimuthal symmetry,}
\begin{equation}
f(\tau_0,x,y,p_x,p_y)
=\rho_T(\tau_0,x,y)\,g_0(p_x,p_y),
\end{equation}
where the momentum distribution satisfies
\[
\int dp_x\,dp_y\, g_0(p_x,p_y)=1.
\]
\tred{Here $g_0(p_x,p_y)$ is obtained from the normalized FONLL spectrum $dN/dp_T$ by converting the radial transverse-momentum distribution into a two-dimensional momentum distribution in the $(p_x,p_y)$ plane. }

\section{PINN for the diffusion equation}

The foundational principle of PINNs is the integration of mathematical models, typically expressed as PDEs along with their corresponding initial and boundary conditions, directly into the neural network's loss function as prior physical knowledge. By embedding these governing equations, PINNs significantly reduce the dependency on large, high-resolution labeled datasets. Consequently, training a PINN is formulated as an optimization problem: the objective is to identify an optimal set of network parameters that minimizes the composite loss function. This process yields a continuous solution that satisfies both the underlying physical laws and the specific constraints of the problem.

\begin{figure*}[!hbt]
\centering
\includegraphics[width=1.05\textwidth]{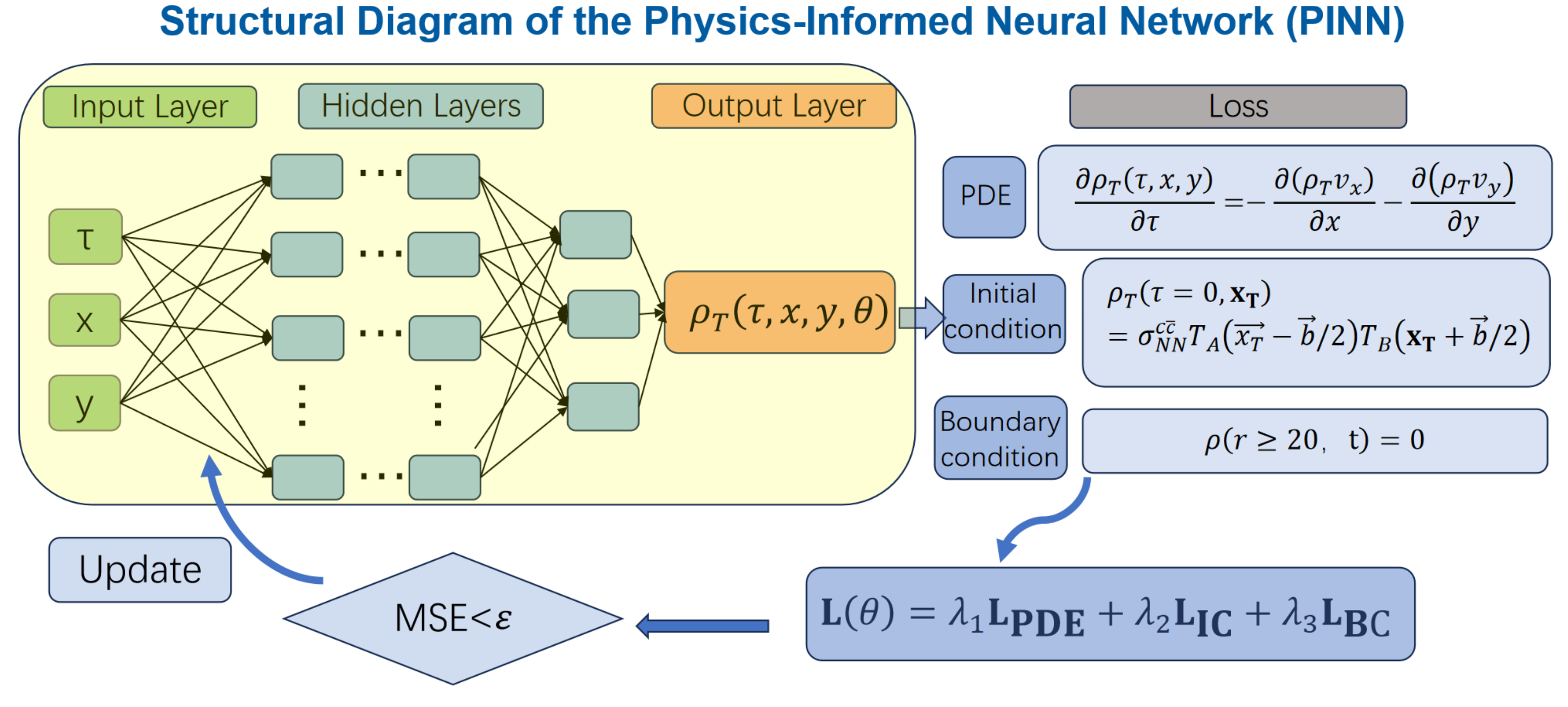}
\centering
\caption{\RaggedRight Schematic architecture of the PINN utilized to resolve the charm quark density $\rho(\tau, x, y)$. The model incorporates the governing diffusion equation, initial condition and boundary conditions as physical constraints.}
\hspace{-0.1mm}
\label{lab-fig-PINN}
\end{figure*}

The input to the PINN consists of the coordinates $(\tau, x, y)$, while the output is the charm quark density $\rho_{nn}(\tau, x, y; \theta)$, where $\theta$ denotes the trainable parameters (weights and biases) of the network. Typically structured as a multilayer perceptron, the network provides an implicit solution representation that is continuous and infinitely differentiable throughout the computational domain. 
\tred{The trainable parameters $\theta$ are iteratively updated to minimize a composite loss function, ensuring that the solution satisfies both the physical laws and the imposed constraints. The hydrodynamic velocity fields $v_x(\tau,x,y)$ and $v_y(\tau,x,y)$ enter the governing equation as known coefficient functions.}

The loss function is defined as a weighted sum of the PDE residual loss $\mathcal{L}_{\text{PDE}}$, the initial condition loss $\mathcal{L}_{\text{IC}}$, and the boundary condition loss $\mathcal{L}_{\text{BC}}$. These components quantify the discrepancy between the network’s prediction, $\rho_{nn}(\tau, x, y; \theta)$, and the target solution at a set of discrete collocation points $(\tau_i, x_i, y_i)$. Specifically, the PDE residual is evaluated by substituting the predicted density into the governing diffusion equation at $N_r$ points randomly sampled within the $(\tau, x, y)$ domain,
\begin{align}
    \mathcal{L}_{\rm PDE} = \frac{1}{N_r} \sum_{i=1}^{N_r} & \left| \frac{\partial \rho_{nn}(\tau_i, x_i, y_i;\mathbf{\theta})}{\partial \tau} \right. \nonumber \\
     &+\frac{\partial [\rho_{nn}(\tau_i, x_i, y_i;\mathbf{\theta})v_x(\tau_i, x_i, y_i)]}{\partial x}  \nonumber \\
     &\left.
    +\frac{\partial [\rho_{nn}(\tau_i, x_i, y_i;\mathbf{\theta})v_y(\tau_i, x_i, y_i)]}{\partial y}\right|^2
\end{align}
Partial derivatives with respect to the coordinates are computed via the network's automatic differentiation capability. To ensure the physical validity of the model, the predicted density $\rho_{nn}(\tau, x, y; \theta)$ is constrained by the prescribed initial and boundary conditions of the PDE. The corresponding initial condition loss $\mathcal{L}_{\text{IC}}$ and boundary condition loss $\mathcal{L}_{\text{BC}}$ are evaluated similarly to the PDE residual, ensuring that the neural network honors all specified physical constraints.
\begin{align}
\label{lab-ic}
    \mathcal{L}_{\rm IC}= &\frac{1}{N_{ic}} \sum_{i=1}^{N_{ic}} |\rho_{nn}(\tau_0, \boldsymbol{ r}_{ic}^i;\mathbf{\theta})-\rho_0(\tau_0, \boldsymbol{r}_{ic}^i)|^2 \\
    \label{lab-bc}
    \mathcal{L}_{\rm BC}=& \frac{1}{N_{bc}} \sum_{i=1}^{N_{bc}} |\rho_{nn}(\tau_i, \boldsymbol{r}_{bc}^i;\mathbf{\theta})-\rho_{bc}(\tau_i, \boldsymbol{r}_{bc}^i)|^2
\end{align}
where $N_{ic}$ denotes the number of collocation points randomly sampled at the initial time $\tau = \tau_0$, where $\boldsymbol{r}_{ic}^i$ represents the transverse coordinates $(x_i, y_i)$ within the physical domain. The term $\rho_0(\tau_0, x, y)$ refers to the initial charm quark distribution. Similarly, $N_{bc}$ indicates the number of temporal points $\tau_i$ sampled along the domain boundaries. These boundaries are defined by a rectangular transverse plane with vertices at $(\pm 20, \pm 20)$ fm. 
The residuals in Eq.~(\ref{lab-bc}) are evaluated at collocation points sampled on all such boundary segments.

The total loss function is a weighted sum of the above terms, 
\begin{align}
\label{lab-eq-loss}
    \mathcal{L}(\theta) = \lambda_1 \mathcal{L}_{\rm PDE}
    +\lambda_2 \mathcal{L}_{\rm IC}
    +\lambda_3 \mathcal{L}_{\rm BC}
\end{align}
\tred{where the hyperparameters $\lambda_1$, $\lambda_2$, and $\lambda_3$ represent the weights of the respective loss terms, ensuring training stability and convergence. These hyperparameters are dynamically adjusted during iterations, and their specific values are detailed in the subsequent sections. Furthermore, during the minimization of $\mathcal{L}(\theta)$, the learning rate is varied across training iterations to enhance the accuracy of the PINN. Compared with a standard supervised neural network, the present framework directly incorporates the governing diffusion equation together with the initial and boundary conditions into the loss function, and therefore does not rely on a large labeled dataset of precomputed numerical solutions.}

As illustrated in Fig. \ref{lab-fig-PINN}, the deep neural network is trained to satisfy the governing PDE while adhering to the prescribed initial and boundary conditions. The architecture consists of four fully connected hidden layers, each containing 256 neurons with Tanh activation functions. To ensure the non-negativity of the predicted density, a Softplus activation is applied to the output layer. Additionally, the velocity field required for the PDE is derived from a hydrodynamic model and pre-processed via a dedicated Velocity Field Cache module. In training the PINN, we adopt a hybrid optimization strategy consisting of a sequential combination of Adam and L-BFGS.
Initially, the Adam optimizer, a first-order stochastic gradient-based method, is employed. Adam’s adaptive learning rate allows it to rapidly escape suboptimal local minima and push the loss towards a favorable basin of attraction during the early training stages. However, due to its stochastic nature, Adam often exhibits oscillatory behavior and slow convergence near the global optimum.
To refine the solution, the training transitions to L-BFGS, a second-order quasi-Newton method. L-BFGS exploits the second-order curvature information of the loss function, enabling a much faster, deterministic convergence with superior numerical precision. This transition ensures that the physical residuals (PDE constraints) are minimized to a tolerance that first-order methods typically fail to reach, thereby significantly enhancing the fidelity and accuracy of the PINN solution.The parameters are updated via backpropagation by minimizing the total loss function $\mathcal{L}(\mathbf{\theta})$, as defined in Eq.~(\ref{lab-eq-loss}).

\section{Numerical results for the diffusion equation}

\begin{figure}[!hbt]
\centering
\includegraphics[width=0.5\textwidth]{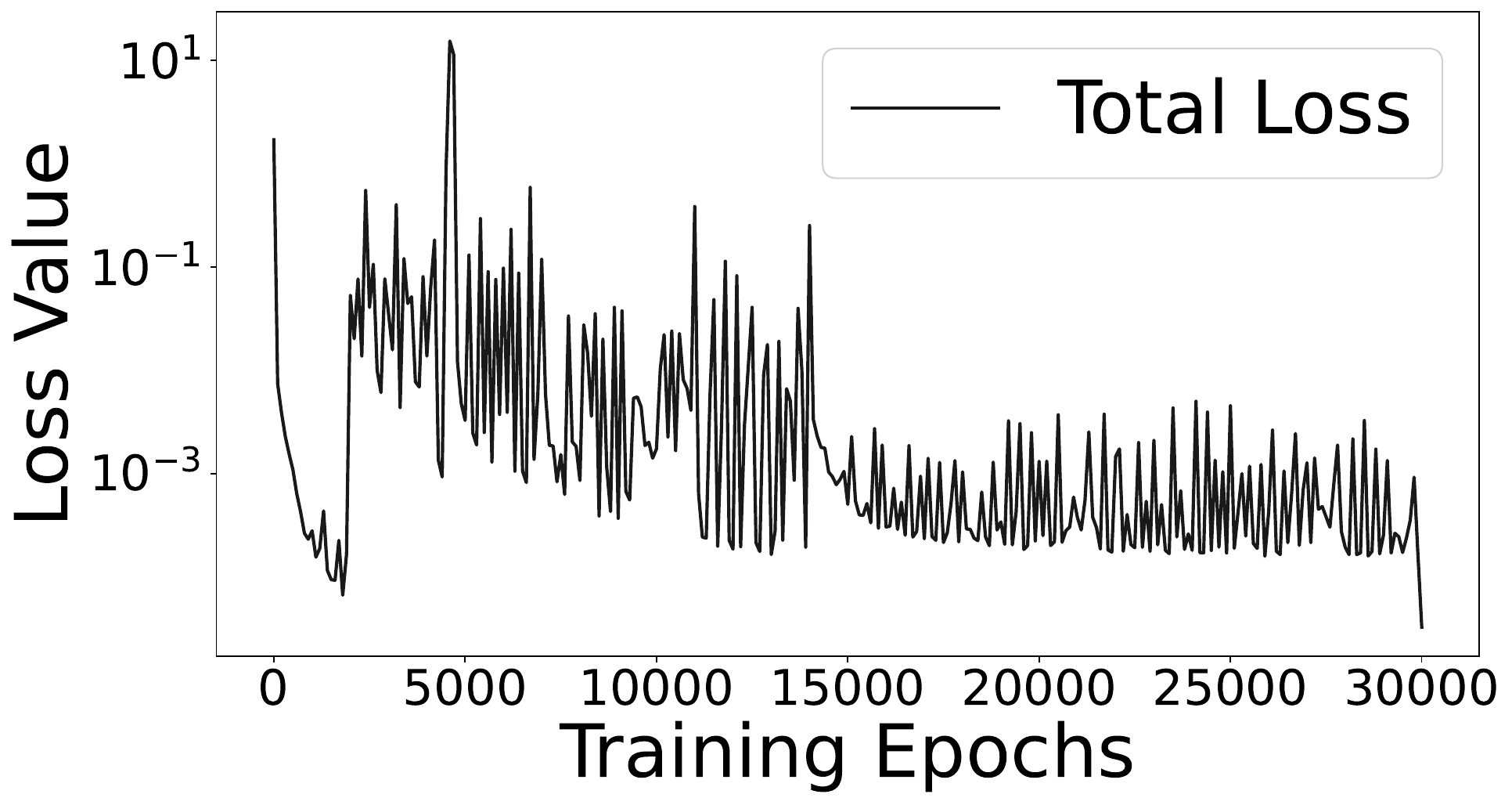}
\centering
\caption{\RaggedRight The total loss function $\mathcal{L}(\theta)$ of the PINNs varying with the number of training epochs. }
\hspace{-0.1mm}
\label{lab-fig-loss}
\end{figure}

The PINN training process iteratively updates the network parameters to minimize the composite loss function, ensuring that the output, $\rho_{\rm PINN}(\tau, x, y)$, satisfies the three physical constraints simultaneously. The convergence history of this loss is presented in Fig. \ref{lab-fig-loss}. During each training epoch, collocation points are sampled over the input domain, with $N_{\tau}$, $N_x$, and $N_y$ denoting the numbers of points used along the temporal and spatial directions, respectively. To enhance the network’s ability to resolve the PDE dynamics, a temporal domain decomposition strategy was employed. The training time interval is extended sequentially through the ranges $[0, 3]$, $[0, 6]$, and $[0, 10]$ fm/$c$ (corresponding to $\tau-\tau_0$ in Table~\ref{tab:training_process}), allowing the PINN to be trained progressively over these stages. The specific learning rates and hyperparameters ($\lambda_1, \lambda_2, \lambda_3$) used to weight the PDE, initial condition (IC), and boundary condition (BC) constraints for each interval are summarized in Table \ref{tab:training_process}. The model was initially trained for $14,000$ epochs using the Adam optimizer, followed by the use of the L-BFGS optimizer to further refine the solution and accelerate convergence toward a deeper local minimum. As illustrated in Fig. \ref{lab-fig-loss}, the total loss function successfully converged to below $10^{-3}$ after a total of $20,000$ epochs.

\begin{table*}[t] 
    \centering
    \large
    \renewcommand{\arraystretch}{1.4} 
    \setlength{\tabcolsep}{6pt}  
    
   \begin{tabular}{|c|c|c|c|c|c|c|c|}
    \hline    

    \multicolumn{5}{|c|}{\textbf{Training process}} & 
    \multicolumn{3}{c|}{\textbf{$\text{L} = \lambda_1 \text{L}_{\text{PDE}} + \lambda_2 \text{L}_{\text{IC}} + \lambda_3 \text{L}_{\text{BC}}$}} \\ 
    \hline    

    \multicolumn{8}{|c|}{\textbf{IC Pretrain}} \\ 
    \hline
    
    \multicolumn{2}{|c|}{Epochs} & \multicolumn{2}{c|}{Learning Rate} & Optimizer & \makebox[1.2cm]{$\boldsymbol{\lambda_1}$} & \makebox[1.2cm]{$\boldsymbol{\lambda_2}$} & \makebox[1.2cm]{$\boldsymbol{\lambda_3}$}\\ 
    \hline
    \multicolumn{2}{|c|}{2000} & \multicolumn{2}{c|}{1e-3} & Adam & 0.0 & 1.0 & 0.0 \\ 
    \hline
    
    \multicolumn{8}{|c|}{\textbf{PDE Curriculum Learning}} \\ 
    \hline
    
    Stage & \begin{tabular}{@{}c@{}}Time range of \\ data\end{tabular} & Epochs & \begin{tabular}{@{}c@{}}Learning \\ Rate\end{tabular} & Optimizer & & & \\ 
    \cline{1-5} 
    
    1 & 0$\sim$3.0 & 5000 & 1e-3 & Adam & & & \\ 
    \cline{1-5}

    2 & 0$\sim$6.0 & 7000 & 5e-4 & Adam & \multirow{-3}{*}{1.0} & \multirow{-3}{*}{10.0} & \multirow{-3}{*}{2.0} \\ 
    \cline{1-5}
    3 & 0$\sim$10.0 & 14000 & 1e-4 & Adam & & & \\ 
    \hline
    
    \multicolumn{8}{|c|}{\textbf{L-BFGS Finetune}} \\ 
    \hline
    
    \multicolumn{3}{|c|}{Learning Rate} & \multicolumn{2}{c|}{Optimizer} & \multirow{2}{*}{5.0} & \multirow{2}{*}{5.0} & \multirow{2}{*}{0.0} \\ 
    \cline{1-5}
    \multicolumn{3}{|c|}{0.8} & \multicolumn{2}{c|}{L-BFGS} & & & \\ 
    \hline
    
    \end{tabular}
    \caption{Training process parameters and loss weights across different stages.}
    \label{tab:training_process}
\end{table*}

\begin{figure*}[hbt]
\centering
\includegraphics[width=0.9\textwidth]{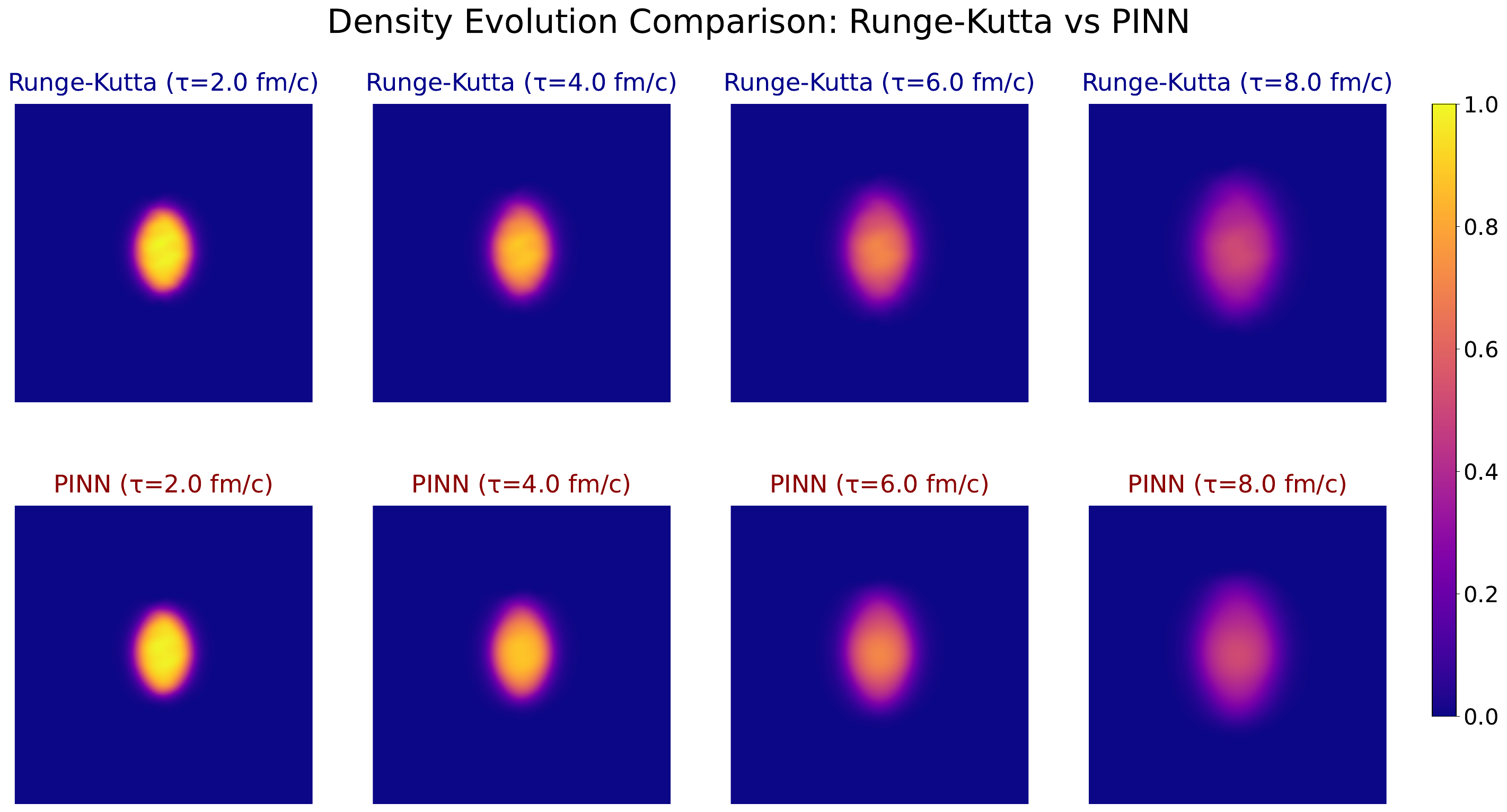}
\centering
\caption{ Charm quark density $\rho_{T}(\tau, x, y)$ at mid-rapidity in 5.02 TeV Pb-Pb collisions with an impact parameter $b=6.0$ fm. The upper and lower panels display the results obtained from the fourth-order Runge-Kutta (RK4) method and the PINN framework, respectively. The color scale indicates the local density magnitude.} 
\hspace{-0.1mm}
\label{lab-fig-den}
\end{figure*}

\begin{figure}[!hbt]
\centering
\includegraphics[width=0.5\textwidth]{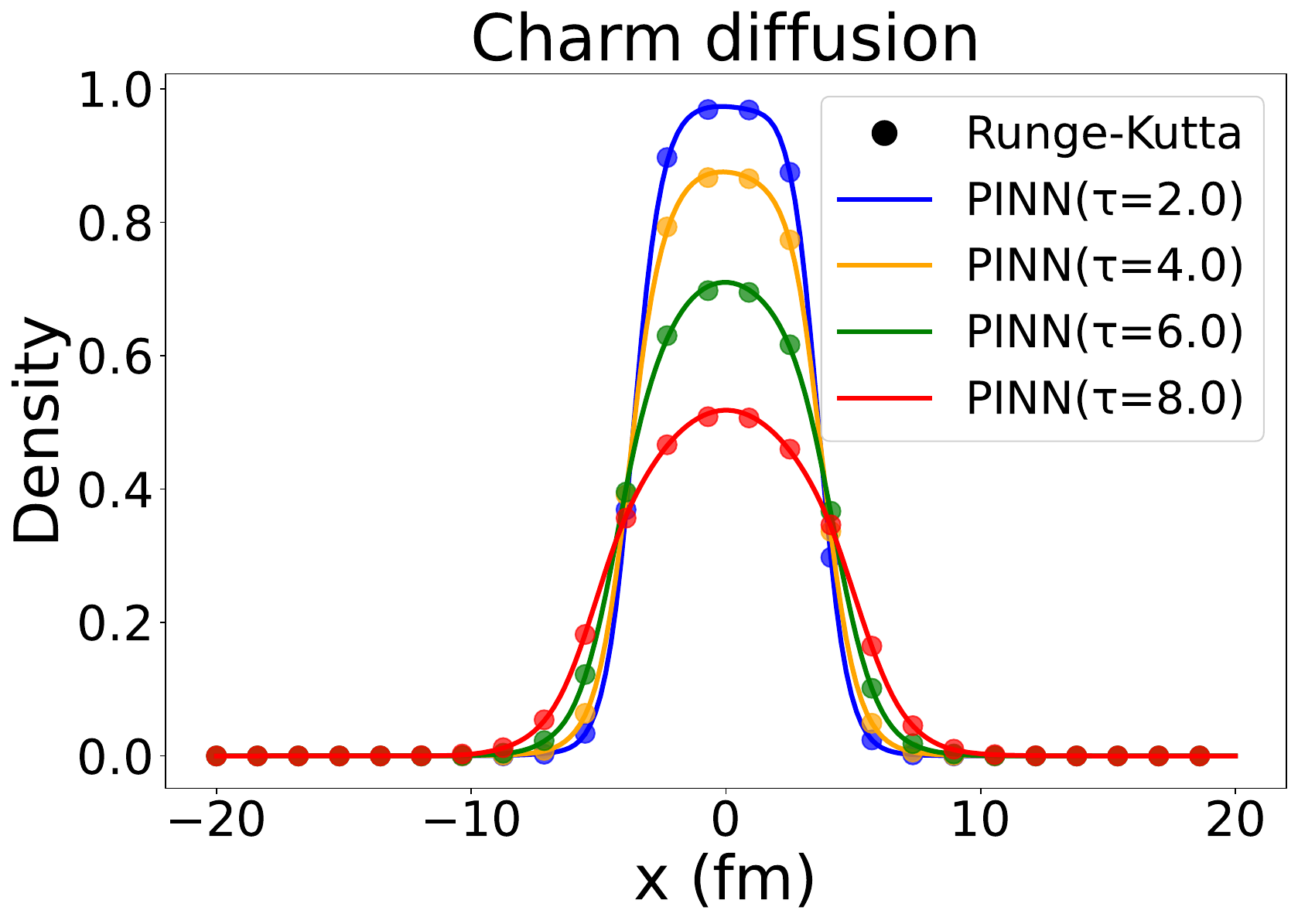}
\caption{\RaggedRight Temporal evolution of the charm quark spatial density \mbox{$\rho_{T}(\tau, x, y=0)$} at mid-rapidity in 5.02 TeV Pb-Pb collisions with an impact parameter \mbox{$b=6.0$} fm. Numerical solutions obtained via the fourth-order Runge-Kutta (RK4) method are represented by discrete points, while the PINN predictions are shown as solid lines.}
\hspace{-0.1mm}
\label{lab-fig-rho-y0}
\end{figure}

\tred{While a low loss function indicates that the PINN output generally satisfies the PDE and its associated initial and boundary conditions, it does not inherently guarantee local accuracy. Therefore, it is essential to evaluate the local residuals at each spatio-temporal point $(\tau, x, y)$ to determine the magnitude of deviation and verify that the model’s predictions remain physically consistent across the entire domain. To establish a baseline for comparison, a fourth-order Runge-Kutta (RK4) method was employed to solve the PDE numerically. By integrating this algorithm with the prescribed initial and boundary conditions, we obtained reference solutions for assessing the accuracy and performance of the PINN framework. Figure~\ref{lab-fig-den} presents the transverse spatial distribution $\rho_T(\tau, x, y)$ of the charm-quark density in the hydrodynamic medium of 5.02 TeV Pb-Pb collisions with an impact parameter $b = 6.0$ fm. The upper and lower panels show the results obtained from the RK4 method and the PINN framework, respectively, with the color scale representing the density magnitude. As shown in the figure, the charm-quark distribution predicted by the PINN is smooth and in close agreement with the RK4 result. For a more quantitative comparison, Fig.~\ref{lab-fig-rho-y0} displays the charm-quark density on the $y = 0$ plane, namely $\rho_T(\tau, x, y=0)$. The RK4 results are represented by discrete points, while the PINN predictions are shown as solid lines. The close overlap between the two demonstrates that the PINN model accurately reproduces the charm-quark diffusion process throughout the spacetime domain considered here.}

\begin{figure}[!hbt]
\centering
\includegraphics[width=0.45\textwidth]{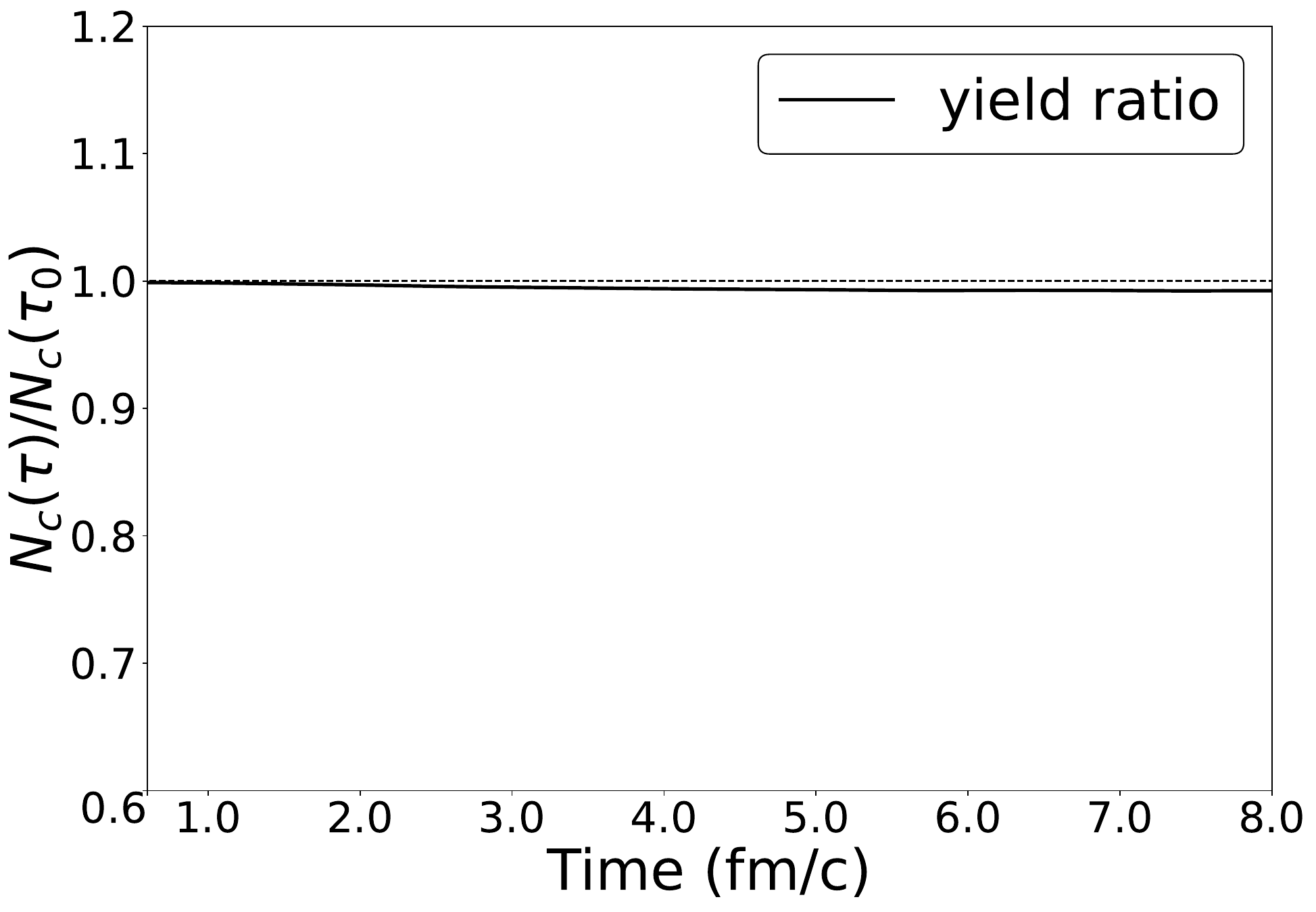}
\centering
\caption{\RaggedRight Temporal evolution of the normalized charm quark abundance, $N_c(\tau)/N_c(\tau_0)$, at mid-rapidity in 5.02 TeV Pb-Pb collisions with an impact parameter $b=6.0$ fm. The total number $N_c$ is obtained by integrating the density $\rho_{\rm PINN}(\tau, x, y)$ predicted by the PINN model.}
\hspace{-0.1mm}
\label{lab-fig-number}
\end{figure}

Since the total number of heavy quarks is conserved within the diffusion equation, one can verify the model's physical consistency by checking whether the PINN output adheres to this constraint. Utilizing the charm quark density $\rho_{\rm PINN}(\tau, x, y)$ predicted by the PINN, the total charm quark abundance at each time step, $N_c(\tau)$, is calculated and presented in Fig. \ref{lab-fig-number}. Across different time intervals, the total quark number remains approximately conserved, exhibiting only negligible deviations. These minor discrepancies are attributed to the inherent numerical approximation errors of the PINN during the optimization process. 
\tred{Overall, the numerical results show that the PINN framework can provide a stable and accurate solution of the charm-quark diffusion equation in a given hydrodynamic background. In the following section, we further examine the applicability of the same methodology to the Fokker-Planck equation, which governs the off-equilibrium dynamical evolution of heavy quarks.}

\section{Extension to the Fokker-Planck equation}

We next apply the PINN framework to the phase-space Fokker-Planck equation introduced above, where the distribution function depends on five variables, $(\tau,x,y,p_x,p_y)$.

For numerical implementation, the phase-space domain is truncated to
$x,y\in[-10,10]~\mathrm{fm}$ and
$p_x,p_y\in[-12,12]~\mathrm{GeV}$.
This reduced domain is adopted to keep the five-dimensional calculation computationally tractable while retaining the dominant support of the distribution.

\subsection{PINN formulation for the Fokker-Planck equation}

Similar to the case of the diffusion equation, the input to the PINN is the phase-space coordinate $(\tau,x,y,p_x,p_y)$, and the network output is the corresponding charm-quark distribution function. Fourier-feature embeddings are introduced for all input coordinates to improve the expressive capacity of the network.

Instead of directly approximating the full distribution function, we adopt a hard-embedded initial-condition construction, following the general trial-solution strategy used in neural-network-based PDE solvers and later PINN formulations~\cite{Lagaris1998,Raissi2019}

\begin{align}
f(\tau,x,y,p_x,p_y)
=
f_0(x,y,p_x,p_y)\,
\exp\!\Big[u(\tau)\,h_\theta(\tau,x,y,p_x,p_y)\Big],
\label{eq:fp_ansatz}
\end{align}
where $f_0(x,y,p_x,p_y)$ denotes the prescribed initial phase-space distribution, $h_\theta$ is represented by the neural network, and $u(\tau)$ is a time-gating function satisfying $u(\tau_0)=0$. In the present work we choose
\begin{align}
u(\tau)=1-\exp[-\lambda(\tau-\tau_0)].
\label{eq:fp_gate}
\end{align}
With this construction, the initial condition is satisfied exactly at $\tau=\tau_0$, while the positivity of the distribution function is automatically ensured for nonnegative $f_0$.

The network parameters are determined by minimizing a composite loss function,
\begin{align}
\mathcal{L}(\theta)
=
\lambda_{\rm PDE}\mathcal{L}_{\rm PDE}
+\lambda_{\rm PHY}\mathcal{L}_{\rm PHY}
+\lambda_{\rm SPEC}\mathcal{L}_{\rm SPEC}.
\label{eq:fp_total_loss}
\end{align}
Here $\mathcal{L}_{\rm PDE}$ is the Fokker-Planck residual loss, constructed from the squared PDE residual evaluated at collocation points in the five-dimensional phase-space domain,
\begin{align}
\mathcal{L}_{\rm PDE}
=
\frac{1}{N_r}\sum_{i=1}^{N_r}
\big|
\mathcal{R}[f](\tau_i,x_i,y_i,p_{x,i},p_{y,i})
\big|^2,
\label{eq:fp_loss_pde}
\end{align}
where $\mathcal{R}[f]$ denotes the residual of the Fokker-Planck equation and all derivatives are computed through automatic differentiation.

The term $\mathcal{L}_{\rm PHY}$ collects auxiliary constraints introduced to improve training stability and to suppress manifestly unphysical behavior caused by the finite computational domain and optimization errors. These constraints include the suppression of the distribution near the truncated momentum boundaries, the control of the total-probability drift, and a weak regularization on the overall evolution of the transverse-momentum spectra. The total probability is defined by
\begin{align}
\mathcal{N}(\tau)
=
\int dx\,dy\,dp_x\,dp_y\,f(\tau,x,y,p_x,p_y),
\label{eq:fp_prob}
\end{align}
and is constrained to remain consistent with the normalization of the prescribed initial distribution during the evolution. In addition, weak spectral regularization is imposed at selected proper times and selected $p_T$ points to suppress unphysical late-time growth and to stabilize the high-$p_T$ softening behavior.

The term $\mathcal{L}_{\rm SPEC}$ is a weak initial spectral-consistency term. It is imposed only at the initial time and only at a sparse set of $p_T$ points. Its role is to ensure that the observable $d^2N/(p_Tdp_T)$ derived from the hard-embedded initial distribution remains consistent with the prescribed input spectrum under numerical integration. It is not used to provide target-spectrum supervision for the subsequent time evolution.

In the training procedure, collocation points are sampled from the interior phase-space domain, from the momentum boundaries, and from the points used to evaluate global normalization and spectral constraints. Sobol low-discrepancy sequences are used in the main sampling procedure. The optimization is performed with AdamW. For selected auxiliary constraints, a gradual weight activation is employed to improve the stability of the high-dimensional training.

\subsection{PINN solution of the Fokker-Planck equation}

In this section, we present the PINN solution of the phase-space Fokker-Planck equation and assess its global consistency and spectral evolution in comparison with the results of the Langevin model~\cite{Chen:2021fye,Cho:2011qv}.
\begin{figure}[!htbp]
\centering
\includegraphics[width=1.00\linewidth]{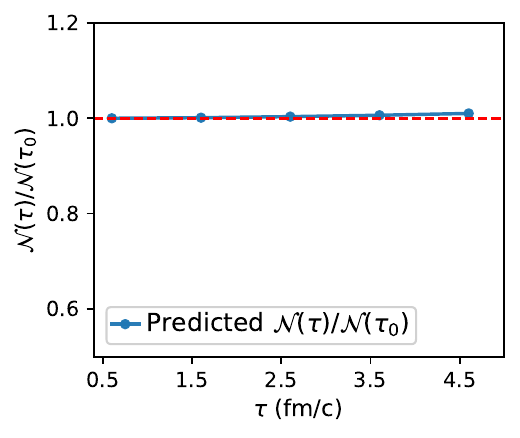}
\caption{\RaggedRight Total probability
$\mathcal{N}(\tau)=\int dx\,dy\,dp_x\,dp_y\,f(\tau,x,y,p_x,p_y)$
obtained from the PINN solution as a function of proper time $\tau$.}
\label{fig:fp_prob}
\end{figure}

Figure~\ref{fig:fp_prob} shows the total probability obtained by integrating the PINN prediction over the truncated phase-space domain at different proper times. The result remains nearly constant throughout the evolution, indicating that the global normalization of the distribution function is preserved with good accuracy and that boundary leakage from the truncated phase-space domain remains small in the present setup.

\begin{figure}[!htbp]
\centering
\includegraphics[width=1.02\linewidth]{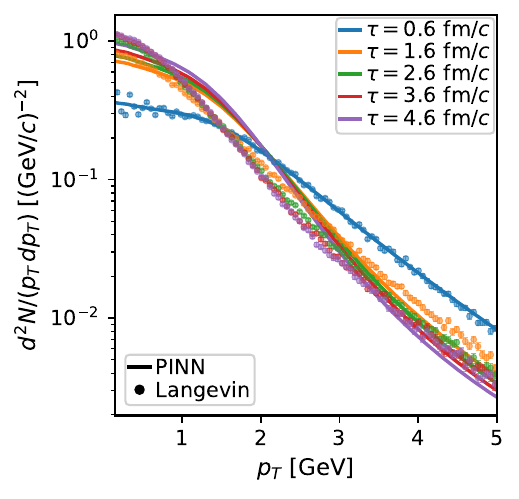}
\caption{\RaggedRight Transverse-momentum spectra $d^2N/(p_Tdp_T)$ at several representative proper times obtained from the PINN solution and compared with the corresponding Langevin results. The dashed curve denotes the initial FONLL-based spectrum at $\tau=\tau_0$. The solid curves show the spectra at different time reconstructed from the learned phase-space distribution, while the markers represent the corresponding Langevin reference results.}
\label{fig:fp_spectrum}
\end{figure}

Figure~\ref{fig:fp_spectrum} shows the transverse-momentum spectra
$d^2N/(p_Tdp_T)$ at several representative proper times obtained from the PINN solution, together with the corresponding the results of Langevin model. The spectrum at the initial time remains consistent with the prescribed FONLL-based input, while at later times the PINN reproduces the overall softening trend observed in the Langevin evolution, including the suppression of the high-$p_T$ tail and the redistribution of strength toward lower and intermediate transverse momenta. This comparison shows that the learned phase-space distribution captures the main spectral features of the Langevin benchmark within the present truncated phase-space setup.

To further quantify the learned solution, we examine several validation diagnostics, including the validation PDE residual, the drift of the total probability during the evolution, and the initial spectral-consistency measure. The corresponding results are summarized in Table~\ref{tab:fp_metrics}. For reference, we also compare the PINN solution with a supervised DNN baseline trained directly on Langevin-generated labels. As shown in Table~\ref{tab:pinn_dnn_compare}, the supervised DNN achieves smaller numerical errors in these metrics, but it relies explicitly on target supervision from precomputed Langevin data. By contrast, the PINN learns the phase-space distribution directly from the transport equation together with physical constraints, without requiring external labels. Its main advantage is therefore methodological: it provides a self-contained physics-informed solver rather than a data-supervised surrogate model.

\begin{table*}[t]
\centering
\large
\setlength{\tabcolsep}{14pt}
\renewcommand{\arraystretch}{1.4}
\caption{Selected validation diagnostics for the PINN solution in the five-dimensional phase-space calculation.}
\label{tab:fp_metrics}
\begin{tabular}{|l|c|}
\hline
\textbf{Diagnostic} & \textbf{Value} \\
\hline
Initial spectral-consistency measure & $5.485\times10^{-3}$ \\
\hline
Validation PDE residual & $1.396\times10^{-5}$ \\
\hline
Probability drift & $6.300\times10^{-3}$ \\
\hline
\end{tabular}
\end{table*}

\begin{table*}[t]
\centering
\large
\setlength{\tabcolsep}{14pt}
\renewcommand{\arraystretch}{1.4}
\caption{Comparison between the PINN solution and the supervised DNN baseline in the five-dimensional phase-space calculation.}
\label{tab:pinn_dnn_compare}
\begin{tabular}{|l|c|c|}
\hline
\textbf{Metric} & \textbf{PINN} & \textbf{Supervised DNN} \\
\hline
Mean PDE residual & $1.396\times10^{-5}$ & $6.874\times10^{-6}$ \\
\hline
Supervision required & No & Yes \\
\hline
Global mass-consistency error & $6.300\times10^{-3}$ & $2.091\times10^{-3}$ \\
\hline
\end{tabular}
\end{table*}

The present Fokker-Planck calculation is intended as a proof-of-principle extension of the PINN framework to a higher-dimensional phase-space transport problem. Within this setup, the numerical results show that the proposed framework can maintain the prescribed initial condition, satisfy the PDE constraint to good accuracy, preserve the total probability, and produce a stable and physically ordered evolution of the transverse-momentum spectrum. The direct comparison with the Langevin benchmark further shows that the PINN captures the main dynamical features of the heavy-quark transport evolution in the expanding medium. Together with the comparison to the supervised DNN baseline, these results indicate that the present framework provides a viable physics-informed solver for heavy-quark transport in a five-dimensional phase-space domain.

\section{Summary}

\tred{In this work, we employ Physics-Informed Neural Networks (PINNs) to explore charm-quark transport in an expanding hot QCD medium, applying the framework to both the coordinate-space diffusion equation and the phase-space Fokker-Planck equation. The numerical results indicate that, within the present setup, the method can provide stable and physically consistent solutions, while also showing potential for treating higher-dimensional transport problems. This study serves as a useful reference and basis for future investigations of non-equilibrium distributions of heavy quarks in the expanding hot QCD medium, which is crucial for understanding the momentum anisotropy of open and hidden heavy-flavor hadrons in heavy-ion collisions.}

\vspace{0.5cm}
{\bf Acknowledgement:}
This work is supported by the
National Natural Science Foundation of China (NSFC)
under Grant Nos. 12575149 and 12175165.

\bibliography{charmflow}
%\bibliography{charmflow}

%\end{spacing}
\end{document}